\documentclass[]{article}

\usepackage{amssymb,latexsym,amsmath}     
\usepackage{authblk}
\usepackage[utf8]{inputenc}
\usepackage{dsfont}
\usepackage[utf8]{inputenc}
\usepackage{graphicx}
\usepackage{mathtools}
\usepackage{amsmath}
\usepackage{subfig}
\DeclareMathOperator{\sgn}{sgn}
\usepackage[colorlinks = true,
linkcolor = blue,
urlcolor  = blue,
citecolor = blue,
anchorcolor = blue]{hyperref}
\usepackage[numbers,square,sort&compress]{natbib}

\usepackage{authblk}
\usepackage{cancel}
\usepackage[noabbrev]{cleveref}
\usepackage{enumitem}

\begin{document}
\title{\textbf{Inverse anisotropic catalysis and complexity}}

\author{ Mojtaba Shahbazi \thanks{mojtaba.shahbazi@modares.ac.ir} }

\author{Mojtaba Shahbazi \thanks{Corresponding author: mojtaba.shahbazi@modares.ac.ir},  Mehdi Sadeghi\thanks{mehdi.sadeghi@abru.ac.ir}\,\,\, \hspace{2mm}\\
	{\small {\em Department of Physics, Faculty of Basic Sciences,}}\\
	{\small {\em Ayatollah Boroujerdi University, Boroujerd, Iran}}
}
\date{\today}
\maketitle

\maketitle
\begin{abstract}
In this work, the effect of anisotropy on computational complexity is considered by CA proposal in holographic two-sided black brane dual of a strongly coupled gauge theory. It is shown that due to the confinement-deconfinement phase transition, there are two different behaviors: with an increase in anisotropy, there is an increase in the complexity growth rate in small anisotropy and a decrease in the complexity growth rate in large anisotropy. In the extreme case, very large anisotropy leads to the unity of the complexity growth rate and the complexity itself, which means that in this case, getting the target state from the reference state is achieved with no effort. Moreover, we suggest that $\frac{1}{M}\frac{dC}{dt}$ is a better representation of system degrees of freedom rather than the complexity growth rate $\frac{dC}{dt}$ and show that how it is related to inverse anisotropic catalysis. In addition, we consider the one-sided black brane dual to the quantum quench and show that increase in anisotropy comes with decrease in complexity regardless of the anisotropy value which is due to the fact that the system does not experience a phase transition.
\end{abstract}

\section{Introduction}
Gauge/gravity duality provides a valuable technique for studying strongly interacting quantum field theories, offering a geometric description of quantum systems \cite{mal}. In this framework, the theory of gravitation resides in the bulk, while the quantum theory exists on the boundary. A dictionary-like correspondence relates quantities in the bulk to those on the boundary. The concept of computational complexity has been introduced \cite{comp} to further investigate the interior of black holes. Computational complexity, a notion from computer science, refers to the minimum number of quantum gates required to transform a reference state into a target state. Several proposals have been made for the holographic dual of computational complexity, such as the "complexity equals volume" (CV) \cite{shock} and "complexity equals action" (CA) \cite{ca} proposals. The CV proposal relates the volume of Einstein-Rosen bridges to computational complexity, while the CA proposal associates the action of a gravitational theory in the Wheeler-DeWitt (WDW) patch to computational complexity. Additionally, a proposal known as CV-2.0  has been put forth, which relates complexity to the spacetime volume restricted to the Wheeler-DeWitt patch \cite{couch}.\\

Recently, an infinite class of surfaces has been proposed as a holographic candidate for complexity \cite{anything}. Interestingly, there appears to be an upper limit for the rate of change of complexity, known as the Lloyd limit \cite{lloyd}:
\begin{equation}
\frac{dC}{dt} \le \frac{2M}{\pi}, \label{lloyd}
\end{equation}
where $M$ is the mass of the black hole. However, the Lloyd limit has been found to be violated in several theories \cite{myers,l1,l2,l3,alish}, leading to modifications that can preserve the limit in some cases, while leaving others still violated \cite{mod1,mod2,mod3}.\\

Anisotropic strongly coupled systems arise in various condensed matter systems under the influence of external fields, such as electric or magnetic fields, as well as in the initial stages of quark-gluon plasmas produced in non-central heavy ion collisions. Strongly coupled anisotropic quantum field theories (QFTs) have been studied holographically using various gravity models, including those with electric/magnetic \cite{e1,e11,e12,e13,e14,e15,e16,e2} and dilatonic/axionic \cite{d1,d11,d12,d13,strong} sources. It has been demonstrated that a non-conformal and anisotropic system with strong interactions corresponds to an Einstein-Dilaton-Axion gravity, where the dilaton field is dual to $Tr F^2$ in the gauge theory, and the axion field $\xi$ is dual to $Tr F \wedge F$ \cite{strong}. Furthermore, it has been shown that the confinement-deconfinement transition temperature decreases with increasing anisotropy, similar to the phenomenon of inverse magnetic catalysis (IMC) observed in lattice QCD \cite{strong}. The gravity model in \cite{strong} describes an uncharged plasma, while the lattice QCD results pertain to a charged plasma, suggesting that the IMC-like effect can occur in uncharged plasmas as well. This idea, termed "inverse anisotropic catalysis" (IAC), has been confirmed in \cite{e2}, where anisotropy alone, without the presence of a magnetic field, leads to the observed effect. However, it is worth noting that IMC does not appear in perturbative QCD, but rather the magnetic catalysis (MC) is observed. There are models that exhibit both IMC and MC for different values of the magnetic field, considering the effect of different angles of the magnetic field and the anisotropy arising due to the pressure gradient in the plasma \cite{interply}.\\

In this paper, we investigate the interplay between anisotropy and complexity in a holographic framework. We consider an Einstein-Dilaton-Axion (EDA) theory to study how anisotropy, characterized by a Lifshitz anisotropic hyperscaling violating exponent, affects the complexity of a two-sided black brane system. This model has been previously explored in the context of holography \cite{EDA1,2,3,EDA2}. Additionally, we explore the Vaidya metric, which is dual to a global quantum quench in the boundary field theory. This allows us to study how closed quantum systems approach equilibrium and how anisotropy influences this process. Our analysis reveals that as the anisotropy increases, the complexity growth rate decreases. This can be attributed to the fact that anisotropy can be induced by the presence of a magnetic field \cite{e1,e11,e12,e13,e14,e15,e16,e2}. The decrease in the complexity growth rate suggests that less "effort" is required to reach the target state from the reference state in the complexity space, i.e., the distance between the states decreases. Furthermore, in the asymptotic regime, where complexity approaches unity at large anisotropy, the system appears to have no difficulty in reaching the target state from the reference state, implying a convergence of the two states. The reduction in the complexity growth rate can be linked to the dilaton field, which governs the coupling of open strings in string theory. Specifically, the increase in anisotropy leads to stronger coupling and, consequently, a lower complexity growth rate. We compare our findings to a previous perturbative analysis \cite{strong}, which also reported a violation of the Lloyd bound at large anisotropy due to the presence of a conformal anomaly. In contrast, the non-perturbative solution considered in this work allows us to study the effect of anisotropy on complexity across the entire range, from small to large anisotropy. Interestingly, we observe two distinct behaviors of anisotropy on complexity, which can be attributed to a confinement-deconfinement phase transition \cite{strong}. Although, the Lloyd bound is respected in small anisotropy, it violates in the large anisotropy. However, in one-sided black brane the Lloyd bound is independent of the effect of anisotropy.\\ 

While the complexity growth rate has been previously associated with the system's degrees of freedom \cite{qcd}, we show that the quantity $\frac{1}{M}\frac{dC}{dt}$\footnote{For a closed system $M$ is constant. It seems that when the mass is changed, the Hilbert space of the system is changed. However, we can consider them as a perturbation and compare the two Hilbert spaces.} is a more appropriate representation, where $M$  is the mass of the system. This quantity is related to the concept of inverse anisotropic catalysis (IAC), which we discuss in detail in the later sections of the paper.\\

The paper is organized as follows: In section \ref{sectwo}, we compute the complexity of the EDA gravity for the two-sided black brane and examine the Lloyd bound. Section \ref{vaidy} focuses on the one-sided black brane case and explores the connection between complexity, degrees of freedom, and IAC. The paper concludes with a summary of the key findings in section \ref{conclusion}.\\

\section{Complexity of two-sided black brane}\label{sectwo}
In this section on shell action in WDW patch is considered for black branes with hyperscaling violating exponent. The gravitational theory could be given by the following action:
\begin{equation}
S=\frac{1}{16\pi G}\int d^{d+2}x \sqrt{-g}\Big(R-\frac{1}{2}(\partial \phi)^2+V(\phi)-\frac{1}{2}Z(\phi)(\partial \xi)^2\Big), \label{action}
\end{equation}
where $Z$ is coupling of dilaton field $\phi$ to axion field $\xi$ and for strongly-coupled gauge theories the coupling and dilaton potential would be \cite{strong}:
\begin{align}
V(\phi)=6e^{\tau \phi},~~Z(\phi)=e^{2\beta \phi}.
\end{align}

A good realization of 5D Einstein-Dilaton-Axion theory can be found in terms of  D3/D7 branes in IIB string theory by $Z(\phi)=e^{2\phi}$ and $V(\phi) = 12$ \cite{string}. The solution of the action \eqref{action} is a Lifshitz anisotropic hyperscaling violation metric which comprises an arbitrary critical exponent $z$ and a hyperscaling violation exponent $\theta$ expressed in terms of constants $\alpha$ and $\beta$. The metric of the dual geometry would be:
\begin{equation}
ds^2=\omega r^{\frac{2\theta}{dz}}\Big(\frac{-f dt^2}{r^2}+\frac{dr^2}{r^2f}+\frac{d\vec{x}^2}{r^2}+\frac{dy^2}{\kappa r^{\frac{2}{z}}}\Big), \label{metric}
\end{equation}
and the function $f$ is given:
\begin{equation}
f=1-\Big(\frac{r}{r_h}\Big)^{d+(\frac{1-\theta}{z})},
\end{equation}
where $r_h$ is the horizon radius and the thermodynamical quantities are as follows:
\begin{align}
&T=\frac{d+\frac{1-\theta}{z}}{4\pi r_h} \label{tem},\\
&M=\frac{V}{16\pi G}\sqrt{\frac{\omega^d}{\kappa}}\frac{dz+1-z-\theta}{z}\frac{1}{r_h^{d+\frac{1-\theta}{z}}}. \label{mass}
\end{align}
The null energy condition (NEC),$T_{\mu \nu} k^{\mu} k^{\nu}>0$, is a fundamental concept in general relativity and quantum field theory that places a constraint on the energy-momentum tensor of a physical system. It requires that for any null vector $k^{\mu}$, the combination of the energy density and pressure in that direction must be non-negative. This ensures that the gravitational field produced by matter/energy is attractive rather than repulsive, helping to stabilize black holes and other compact objects, and limiting the types of exotic matter that can exist in nature. Violations of the NEC, though theoretically possible, are thought to be uncommon.\\
By applying the null energy condition, we have:
\begin{equation}
(z-1)(1+dz-\theta)\ge 0,~~\theta ^2+dz(1-\theta)-d \ge 0.
\end{equation}
One can find eligible ranges for $z$ and $\theta$, that is revealed $z_-\ge z,  z\ge z_+$ and $\theta_-\ge \theta, \theta\ge \theta_+$:
\begin{align}
\theta_{\pm}=\frac{1}{2}\Big(dz\pm \sqrt{(dz)^2-4dz+4d}\Big),~~z_{\pm}=\frac{1}{2d}\Big(d+\theta-1\pm |d-\theta+1|\Big).
\end{align}
In the following we suppose that $d>z $ and $d>\theta$ such that satisfy null energy condition (NEC).

The trace of the equation of motion leads to:
\begin{align}
R-\frac{1}{2}(\partial \phi)^2-\frac{1}{2}e^{2\beta \phi}(\partial \xi)^2+(\frac{d+2}{d})6e^{\tau \phi}=0,
\end{align}
then the density of the Lagrangian reads as:
\begin{equation}
\sqrt{-g}\Big(R-\frac{1}{2}(\partial \phi)^2-\frac{1}{2}e^{2\beta \phi}(\partial \xi)^2+6e^{\tau \phi}\Big)=-\Big(\sqrt{\frac{\omega^{d+2}}{\kappa}}r^{\frac{(d+2)\theta}{dz}-\frac{1}{z}-(d+1)}\Big)\frac{12}{d}e^{\tau \phi}. \label{den}
\end{equation}
For the IR limit one can find exact solution \cite{strong}:
\begin{equation}
\phi=c_1 \log (c_2 r)+c_3, \label{dilaton}
\end{equation}
where $c_2=\omega^{\frac{dz}{2\theta-2dz}}$ that can be called anisotropic parameter, $\kappa=\frac{1}{a}c_2^{\frac{2d-2dz}{dz}}$ and $a,c_1,c_3$ are constant. The right hand side of eq. \eqref{den} would be:
\begin{equation}
-\frac{12}{d}\sqrt{\frac{\omega^{d+2}}{\kappa}}e^{\tau c_3} c_2^{\tau c_1}~ r^{\frac{(d+2)\theta}{dz}-\frac{1}{z}-(d+1)+\tau c_1},
\end{equation}
then eq. \eqref{action} leads to:
\begin{equation}
S=-\frac{3V_d}{4\pi Gd}\sqrt{\frac{\omega^{d+2}}{\kappa}}e^{\tau c_3} c_2^{\tau c_1} \int  dt dr~ r^{\frac{(d+2)\theta}{dz}-\frac{1}{z}-(d+1)+\tau c_1}. \label{reform}
\end{equation}

If one is interested in computing the action in WDW patch should add some terms to the action \cite{pad,myersnul}:
\begin{equation}
\begin{aligned}
S&=\frac{1}{16\pi G}\int d^{d+2}x \sqrt{-g}\Big(R-\frac{1}{2}(\partial \phi)^2+V(\phi)-\frac{1}{2}Z(\phi)(\partial \xi)^2\Big)+\frac{1}{8\pi G}\int d^{d+1}x \sqrt{-h}K \\
& \pm \frac{1}{8\pi G}\int d^dx \sqrt{-\gamma} \log (a_{ij}), \label{wdw}
\end{aligned}
\end{equation}
where for two light-like vectors and for a light-like and a time-like vector respectively there would be: 
\begin{align}
a_{ij}=|\frac{n_i.n_j}{2}|,~~a_{ij}=|n_i.k_j|.
\end{align}
Gibons-Howking-York(GHY) term is read as follows:
\begin{equation}
S_{\text{GHY}}=\frac{\sgn(j)}{8\pi G}\int_{\Sigma}\sqrt{-h}Kd^{d+1}x, \label{ghy}
\end{equation}

\begin{figure}[htp]
    \centering
    \includegraphics[width=8cm]{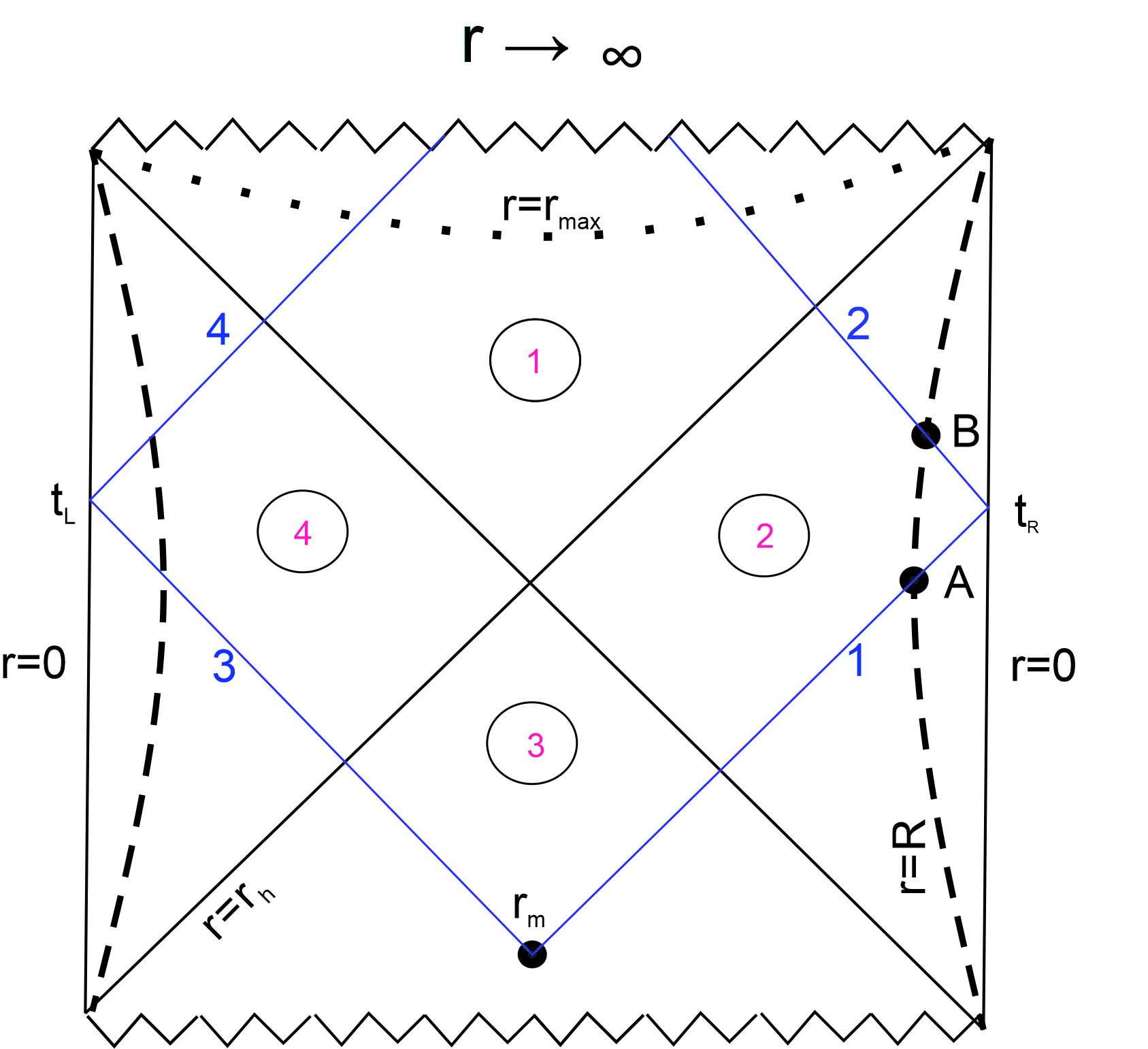}
    \caption{WDW patch of a two sided black brane.} \label{fig:two}
\end{figure}

\begin{figure}[htp]
    \centering
  \subfloat[][]{ \includegraphics[width=15cm]{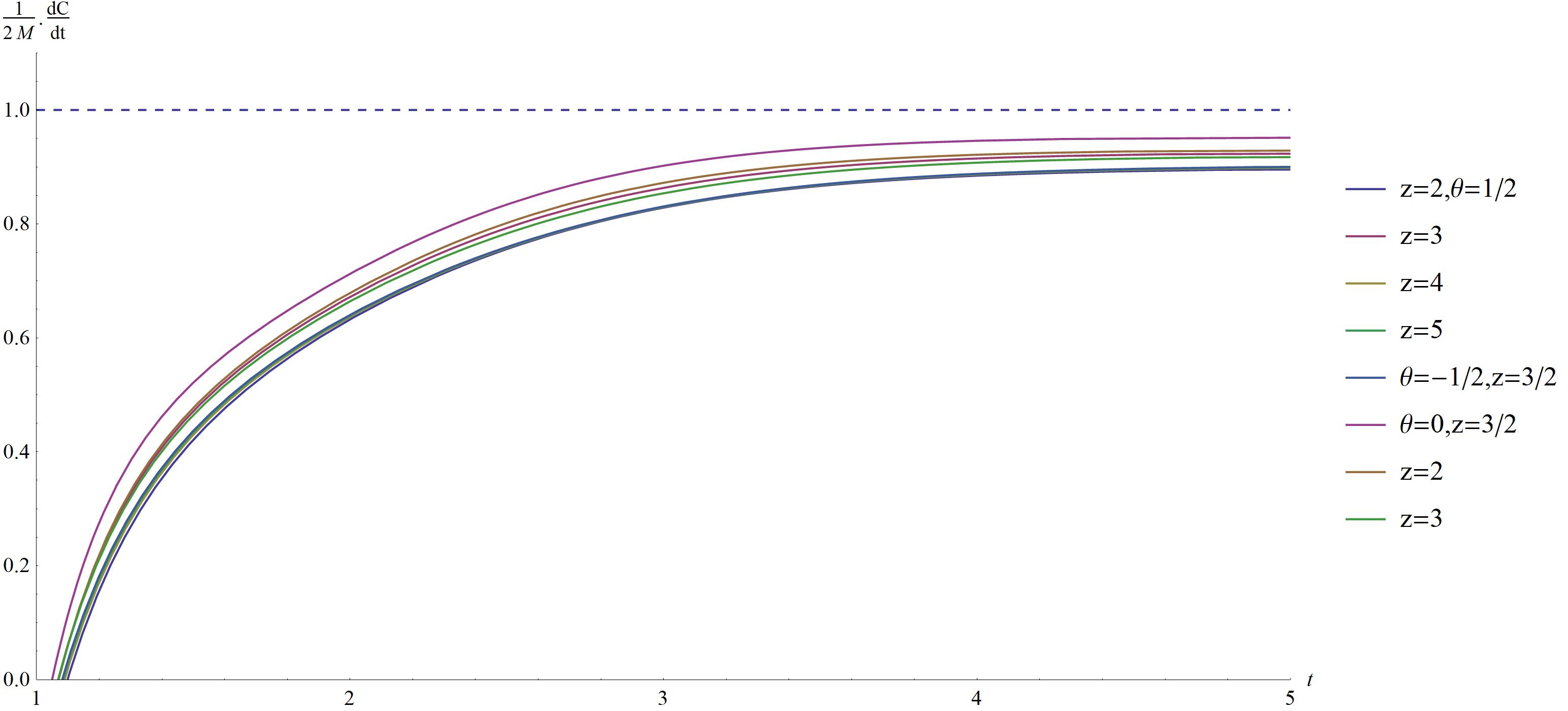}\label{at}}
\vfill
  \subfloat[][]{\includegraphics[width=13cm]{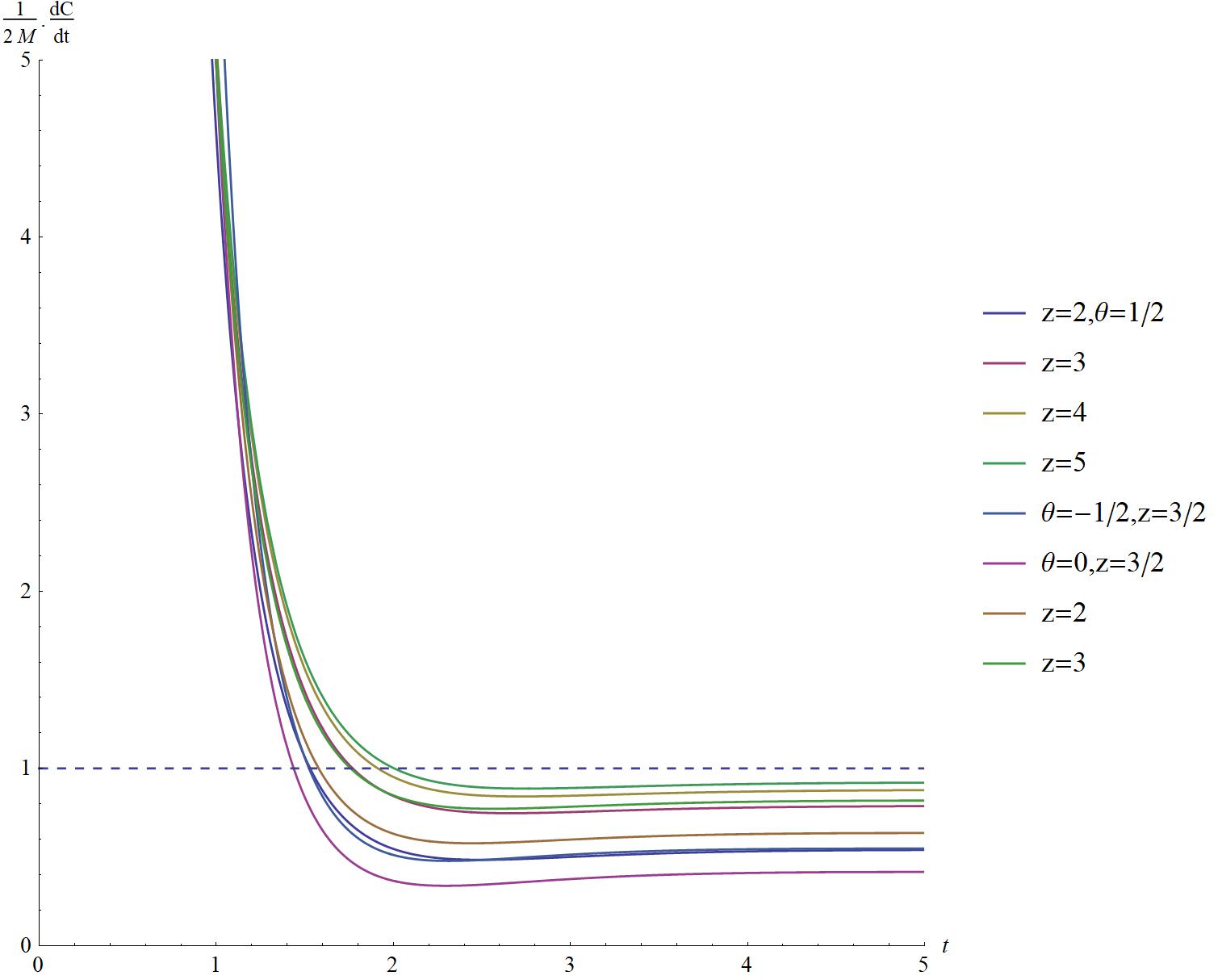}\label{10}}
   
    \caption{Complexity growth rate of the two-sided black brane for different values of $z$ and $\theta$ (a)  $c_2=1$, (b) $c_2=10$.}
\end{figure}

$\sgn(j)$ for time-like surface $\Sigma$ is 1 and for space-like surface if it lies to the past of the bulk is 1 otherwise -1 and for light-like surfaces GHY term vanishes. $K$ is extrinsic curvature of the (d+1)-dimensional hypersurface and $h$ is determinant of the induced metric. 
By CA proposal one can compute the complexity growth rate:
\begin{align}
\frac{dC}{dt}=\frac{1}{\pi}\frac{dS_{\text{total}}}{dt}.
\end{align}
By knowing the fact that $\frac{dr_m}{dt}=\frac{f(r_m)}{2}$, complexity growth rate is given (for details see appendix \ref{detail}):

\begin{equation}
\begin{aligned}
\frac{dS_{\text{total}}}{dt}=&\frac{V_d}{8\pi G}\sqrt{\frac{\omega^d}{\kappa}}\frac{dz+1-\theta-\frac{\theta}{dz}}{z~r_h^{d+(\frac{1-\theta}{z})}}\Bigg(1+\frac{1}{2}\Big(\frac{r_h}{r_m}\Big)^{\frac{dz+1-\theta}{z}}f(r_m)\Big(\frac{\theta-1-dz+z}{dz+1-\theta-\frac{\theta}{dz}}\Big)\Big(\log\omega-\log f(r_m)\Big)\\
&+\frac{6\omega~z ~e^{\tau c_3} ~c_2^{\tau c_1}~r_h^{\tau c_1+\frac{2\theta}{dz}}}{\frac{d}{z}(dz+1-\theta-\frac{\theta}{dz})(\tau c_1-dz-1+\theta+\frac{2\theta}{d})}\Big(\frac{r_h}{r_m}\Big)^{\frac{dz+1-\theta}{z}-\tau c_1-\frac{2\theta}{dz}}\Bigg). \label{anig}
\end{aligned}
\end{equation}
In Fig.(\ref{at}) we set $c_1=c_3=1$. As it is shown, the Lloyd bound is respected in these models for different values of parameters. Moreover, Fig.(\ref{10}) shows the violation of the Lloyd bound in the large anisotropy. The difference between small and large anisotropy effects on complexity arises due to the phase transition of confinement-deconfinement type in the system.
\\

\section{Complexity for Vaidya metric}\label{vaidy}
Vaidya metric describes formation of a black hole which implies the mass of the system increases in time. In addition, the metric could correspond a gravitational description for a global quantum quench in a field theory that means exiting a system out of equilibrium and let it evolve in time. Study of global quenches provides an understanding of how closed quantum systems reach equilibrium. Furthermore, owing to the fact that the complexity grows for a long time after equilibrium, it is expected that considering complexity growth rate for theses systems does not come up with a new insight. Nonetheless, they can provide an examination of the Lloyd bound.

For the action \eqref{action} by adding an infalling null shell, the metric is given by:
\begin{equation}
ds^2=\omega r^{\frac{2\theta}{dz}}\Big(-\frac{f~dv^2}{r^2}-\frac{2dvdr}{r^2}+\frac{d\vec{x}^2}{r^2}+\frac{dy^2}{\kappa r^{\frac{2}{z}}}\Big),
\end{equation}
in which the function $f$ would be:
\begin{figure}[h!]
    \centering
    \includegraphics[width=7.5cm]{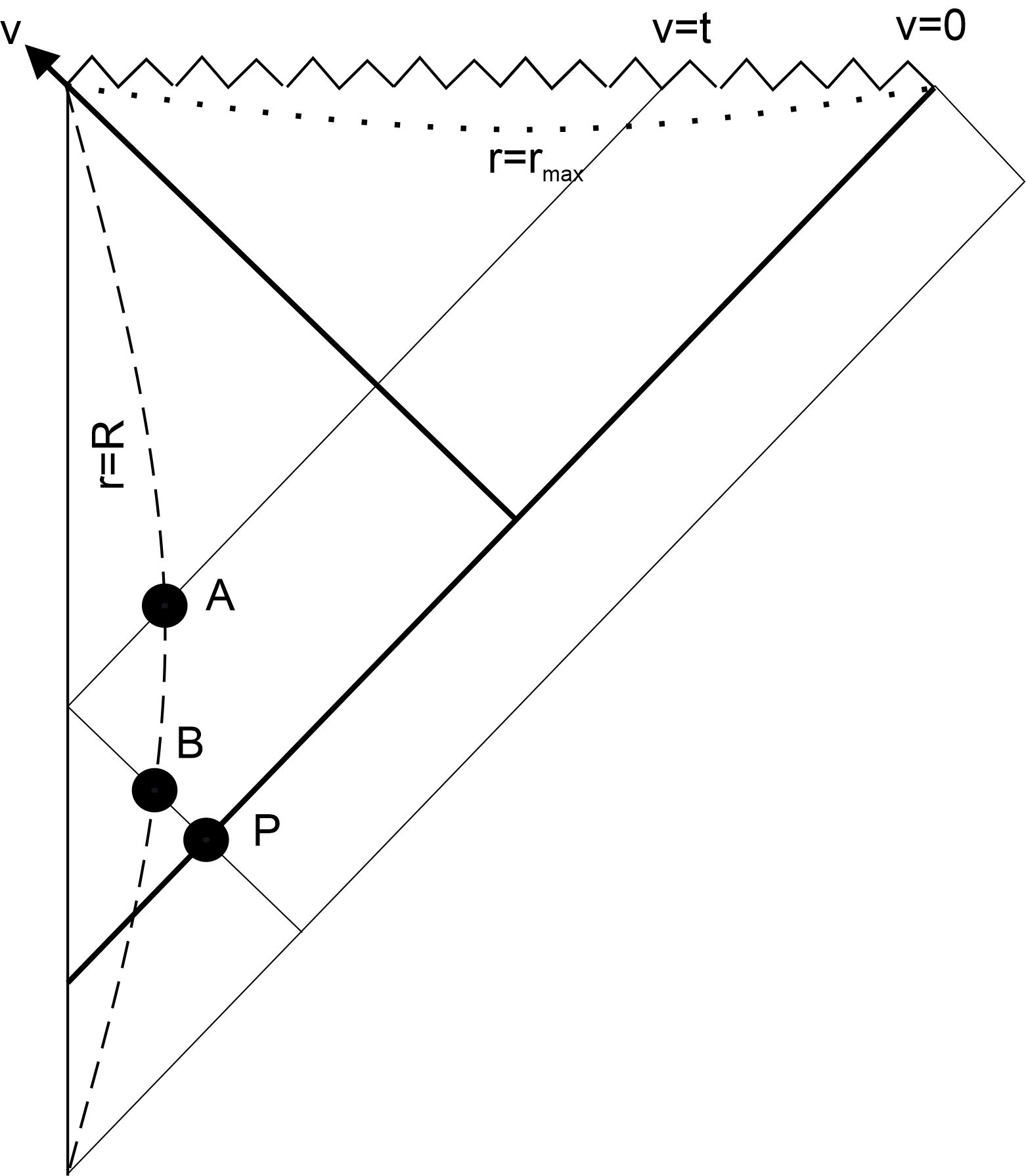}
    \caption{WDW patch of a one sided black brane.}
\end{figure}
\begin{equation}\label{step}
f(r,v)=
    \begin{dcases}
        1 & v<0 \\
       1-\Big(\frac{r}{r_h}\Big)^{d+\frac{1-\theta}{z}} & v>0 \\
           \end{dcases}
\end{equation}
The total action in Vaidya metric would be:
\begin{align}
S_{\text{total}}=S^{v>0}+S^{v<0}.
\end{align}
The complexity growth rate is time derivative of the complexity; moreover, by invoking relation $\frac{dr_P}{dt}=\frac{f(r_P)}{2}$ one has (for details see appendix \ref{vaidyaa}):
\begin{equation}
\begin{aligned}
\frac{dS_{\text{total}}}{dt}=&\frac{V_d}{8\pi G} \sqrt{\frac{\omega^d}{\kappa}}\frac{\Big(\frac{dz+1-\theta-\frac{\theta}{d}}{z}\Big)}{r_h^{\frac{dz+1-\theta}{z}}}\Bigg(1+\frac{1}{2}\Big(\frac{dz-\theta+1-z}{dz-\theta+1-\frac{\theta}{d}}\Big)\Big(\frac{r_h}{r_P}\Big)^{\frac{dz+1-\theta}{z}}f(r_P) \log f(r_P)\Bigg).
\end{aligned} \label{vaidyag}
\end{equation}

\begin{figure}[htp]
    \centering
    \includegraphics[width=16cm]{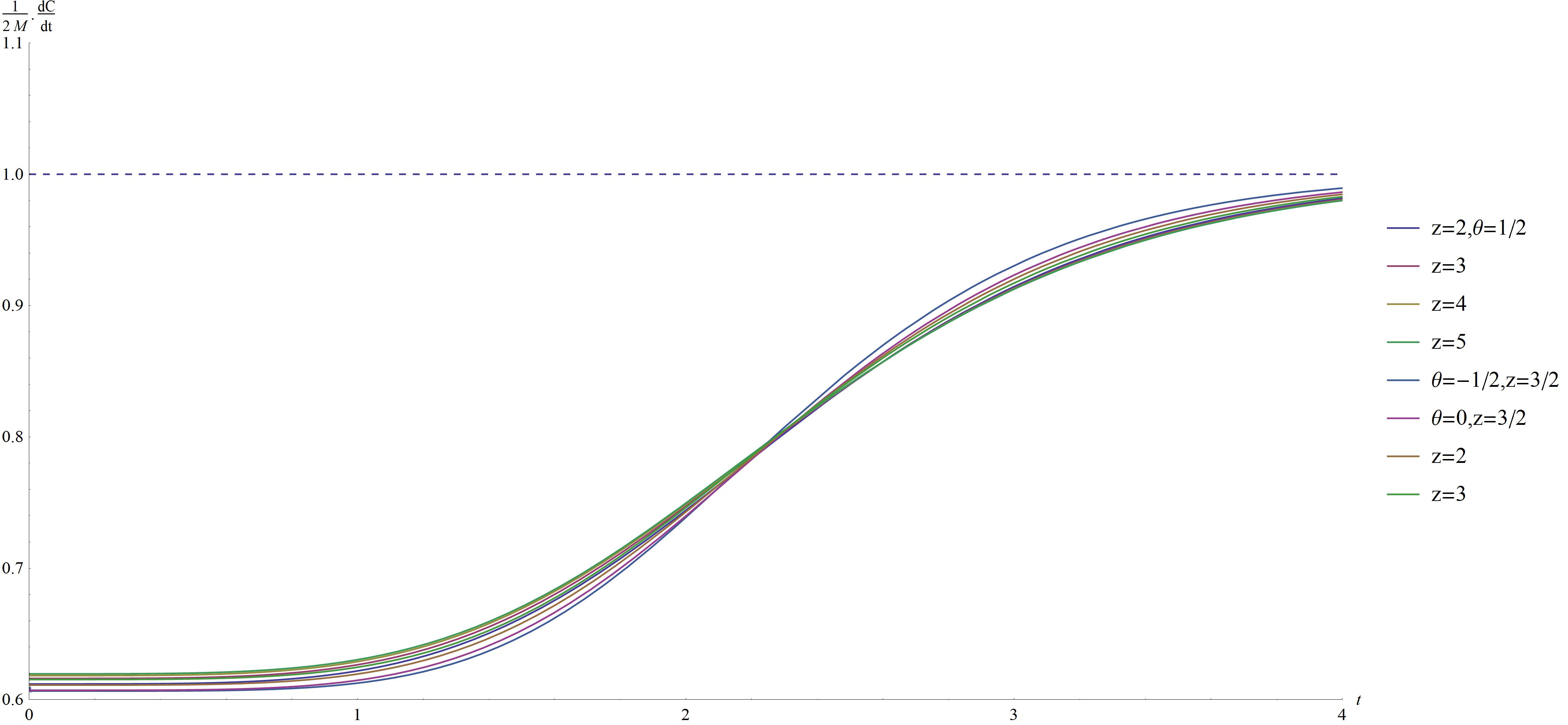}
    \caption{Complexity growth rate of Vaidya metric for different values of $z$ and $\theta$ in terms of the final energy of the system.}\label{vlo}
    \end{figure}

The complexity growth rate in Vaidya solution could be seen in Fig(\ref{vlo}).

Due to the fact that the anisotropy parameter only appears in the mass of the black brane \eqref{mass} in eq. \eqref{vaidyag}, then $\frac{1}{M}\frac{dC}{dt}$\footnote{By $M$ we mean the value that the complexity growth rate saturates at late times. However, one can consider the mass of the black brane in terms of $v$ as $M(v)=\frac{M}{2}\Big(1+tanh(\frac{v}{v_0})\Big)$ where $v_0$ refers to the thickness of the infalling dust and if it goes to zero we get the step function \eqref{step}. Nonetheless, the result gets some correction but the functionality to the anisotropy parameter and time would not change.} on the contrary of the two-sided black brane is independent of the anisotropy, in other words although the complexity growth rate increases with the increase in the anisotropy parameter, the Lloyd bound (complexity growth rate divided by the mass of the system) is independent of the anisotropy parameter and the Lloyd bound is satisfied. The behavior originates from the injection of energy globally in the system which is dual to the formation of the black brane. In this manner the effects of the increase in the degrees of freedom due to the mass of the black brane is dominant to the effects of the anisotropy due to the dilaton field.

\section{Physical reasons behind the complexity and IAC}  \label{phy}
\begin{figure}[htp]
    \centering
    \includegraphics[width=8cm]{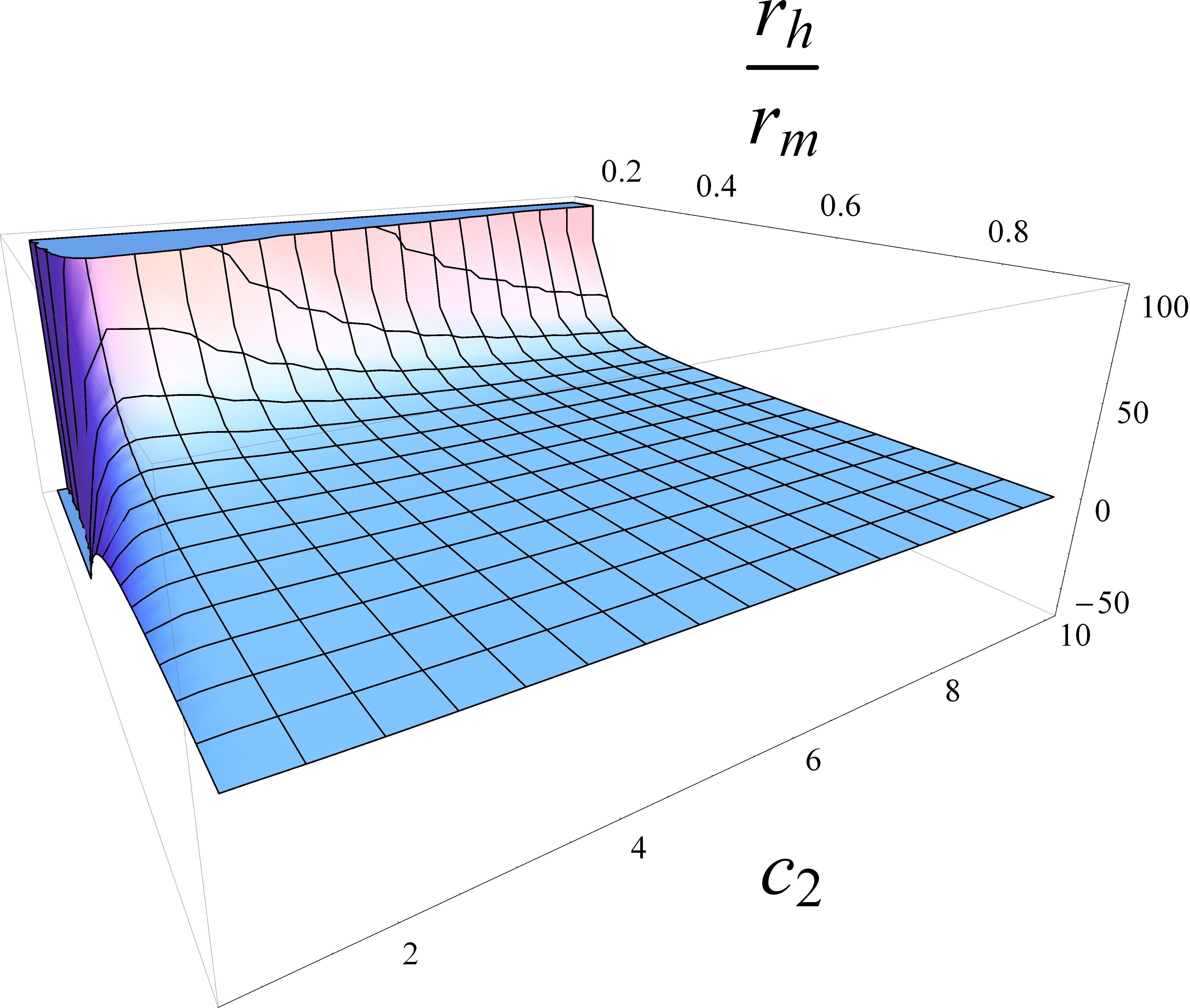}
    \caption{Complexity growth rate density ($\frac{1}{V}\frac{dC}{dt}$) of two-sided black brane for values of $z=2$ and $\theta=1/2$ in terms of anisotropy parameter $c_2$. }\label{ani}
\end{figure}

The effect of the anisotropy on the complexity growth rate for different values of $c_2$ has been shown in Fig(\ref{ani}) and a turning point which could be addressed to the confinement-deconfinement phase transition as reported in \cite{strong}. Fig(\ref{ani}) depicts that complexity growth rate decreases when the anisotropy increases; in addition, the complexity itself decreases as anisotropy increases (see appendix \ref{complexity}):
\begin{align}
S_{\text{total}}\approx c_2^{-d+1\frac{\theta-1}{z}}\log(c_2),
\end{align}
put it another way, target states get closer to the reference state in complexity space. In view of the fact that our solution is non-perturbative, then by very large anisotropy parameter, the complexity growth rate would be unity.

In the following the physical reasons behind the behavior of the complexity growth rate are regarded by the coupling role and considering the system degrees of freedom. 
\subsection{Coupling}
The physical reason behind the overall effect is due to the dilaton field in sting theory that illustrating how open strings are coupled to each other. The more value of $c_2$ (the more coupling of open strings), the more difficulty reaching to the target state \cite{ghodrati}. The same result about effect of dilaton on complexity has been shown in \cite{dil}. However, around $c_2=1$ there is an increase in complexity growth rate. The same behavior appears in  \cite{seyed} where their solution is perturbative of order $O(c_2^2)$. This increase could be addressed to the mass of the black brane \eqref{mass}, in other words, away from $c_2=1$ the mass of the black brane makes large contribution towards the complexity growth rate. Although in our work the solution \eqref{dilaton} is in IR limit, that is not perturbative. Interestingly, in \cite{seyed} for large anisotropy parameter their model displays a conformal anomaly and they expected that  it leads to the violation of the Lloyd bound. Here, we show that the violation of the Lloyd bound for large anisotropy could be seen in Fig (\ref{10}).

The effect of coupling on the complexity could be explained by energy-time uncertainty relation in quantum mechanics. The minimum time it takes the system from an initial state goes to another state is given by:
\begin{align}
\Delta t\sim \frac{\hbar}{\Delta E}.\label{uncer}
\end{align}
In Zeeman (or Stark) effect energy levels of the system are corrected, for example in Hydrogen atom in weak field approximation by \cite{gri}:
\begin{align}
E_n=-\frac{13.6eV}{n^2}+\mu B(m_l+2m_s).\label{energy}
\end{align}
When the energy levels are changed the time it takes the system undergoes the time evolution is changed, then the complexity growth rate is varied\footnote{When the coupling is changed the Hilbert space of the system is varied, then it is convenient to see the complexity is about what states. Moreover, when the coupling makes changes of the complexity growth rate, the question would be: is this difference attributed to different states or gates? It means that due to the different reference and target states the complexity undergoes a change or due to the different gates. By energy-time uncertainty relation \eqref{uncer} we argued that the states are changed so that, say, the ground state in Hilbert space $\mathcal{H}_{B_1}$ with coupling $B_1$ goes to the first excited state in $\mathcal{H}_{B_1}$ and correspondingly the ground state in Hilbert space $\mathcal{H}_{B_2}$ with coupling $B_2$ goes to the first excited state in $\mathcal{H}_{B_2}$. It is true that the two Hilbert spaces are different but there is a correspondence between the states in the unperturbed and perturbed one in quantum perturbation theory. However, in the bulk side and holographic complexity there are ambiguities in quantum gates and the reference state \cite{anything}.}. 
Nonetheless, this is in no contradiction with the Jackiew-Taitelboim (JT) gravity and its dual ensemble of random coupling quantum theories \cite{random}, since the holographic complexity of JT gravity is an averaging of complexity of the  ensemble members \cite{averaging}.

Moreover, Vaidya solutions study the thermalization of systems. \eqref{vaidyag} shows that the complexity growth rate decreases as the anisotropy increases:
\begin{align}
\frac{dS_{\text{total}}}{dt}\propto c_2^{-d+1\frac{\theta-1}{z}}.
\end{align}
As anisotropy grows the coupling of open strings is intensified which means that the number of operators acting on states in time (complexity growth rate) decreases. It is seems that as the anisotropy and the coupling increase the thermalization time decreases \cite{aref} because of the fact that the more strong coupling the more rapid thermal spreading and as a consequence the number of quantum gates (operators) required to turn the reference state to the target state decreases.

\subsection{Degrees of freedom}
One can show that the complexity growth rate at late times is proportional to the number of degrees of freedom in an Einstein-Dilaton gravity \cite{qcd}:
\begin{align}
\frac{1}{T^4}\frac{dC}{dt} \propto N^2,
\end{align}
where $N$ is the group rank of dual CFT and roughly speaking the number of degrees of freedom. If one takes this as an assumption then there would be:
\begin{equation}
\frac{dC}{dt}\propto M \propto N^2.
\end{equation}
The above relation and Fig(\ref{ani}) show that as anisotropy parameter grows, the number of degrees of freedom is reduced. By thermodynamics of black holes we can see that for our black brane:
\begin{align}
\frac{M}{T^4}=\frac{s}{T^3}\propto N^2.
\end{align}
However, this result is in glaring contradiction with Fig(\ref{ani}) where at large anisotropy, the complexity in the deconfinement phase is lower than the confinement phase. Nonetheless, if one omits the effect of mass of the black brane and plots $\frac{1}{M}\frac{dC}{dt}$ in terms of the anisotropy, it makes sense \footnote{It worth mentioning that the energy of the system $M$ is constant for a fixed anisotropy parameter; however, different anisotropy parameters refer to different systems. In this manner, comparing the different systems makes $M$ come into play.}.

\begin{figure}[htp]
    \centering
    \includegraphics[width=8cm]{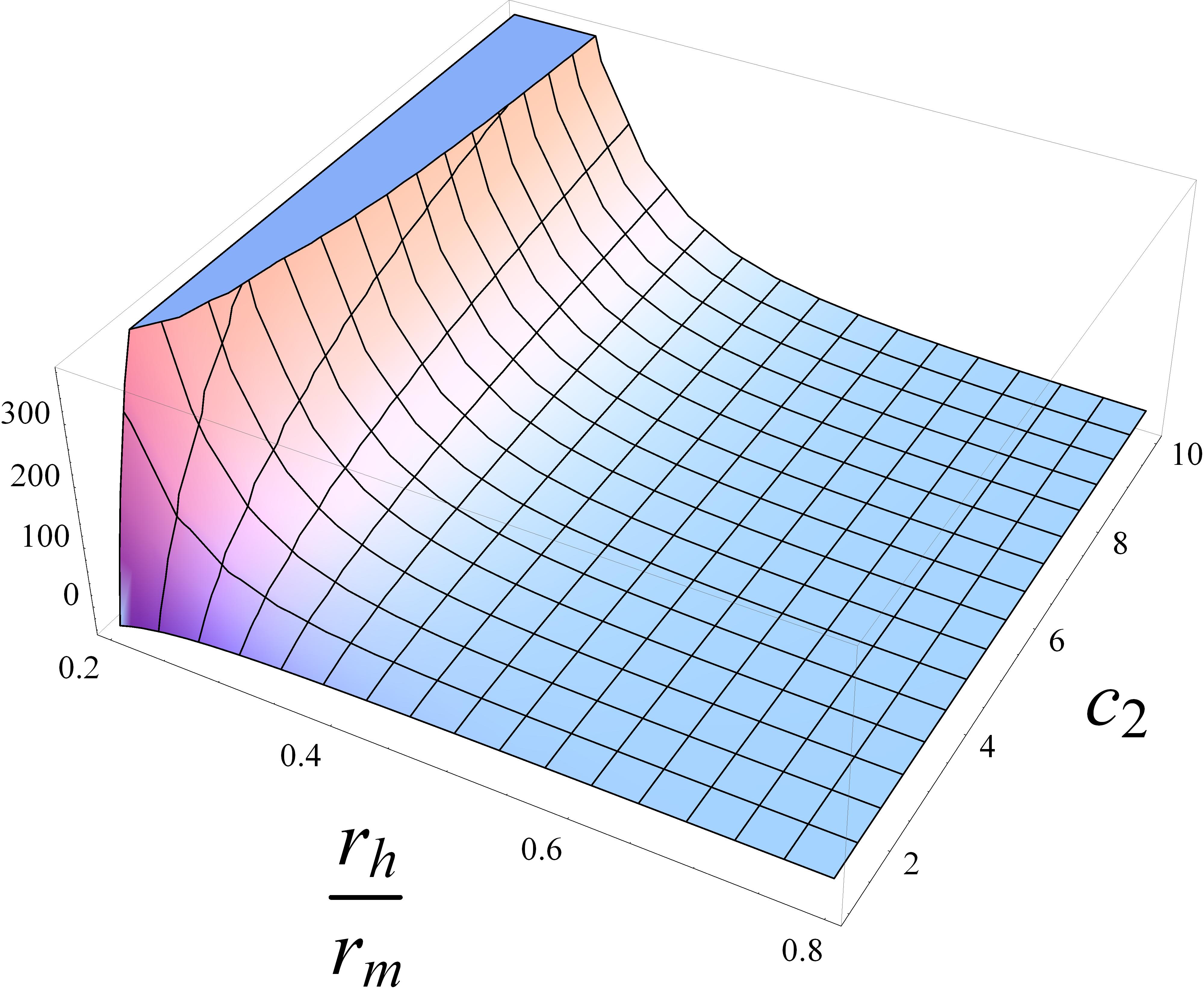}
    \caption{Complexity growth rate divided by the mass of two-sided black brane $\frac{1}{M}\frac{1}{V}\frac{dC}{dt}$ for values of $z=2$ and $\theta=1/2$ in terms of anisotropy parameter $c_2$. }\label{awm}
\end{figure}

As in Fig(\ref{awm}), it is manifest that at large anisotropy where the system is in the deconfinement phase, $\frac{1}{M}\frac{dC}{dt}$ increases. Then, it is suggested that one should interpret $\frac{1}{M}\frac{dC}{dt}$ as the degrees of freedom.

\begin{figure}[htp]
    \centering
    \includegraphics[width=8cm]{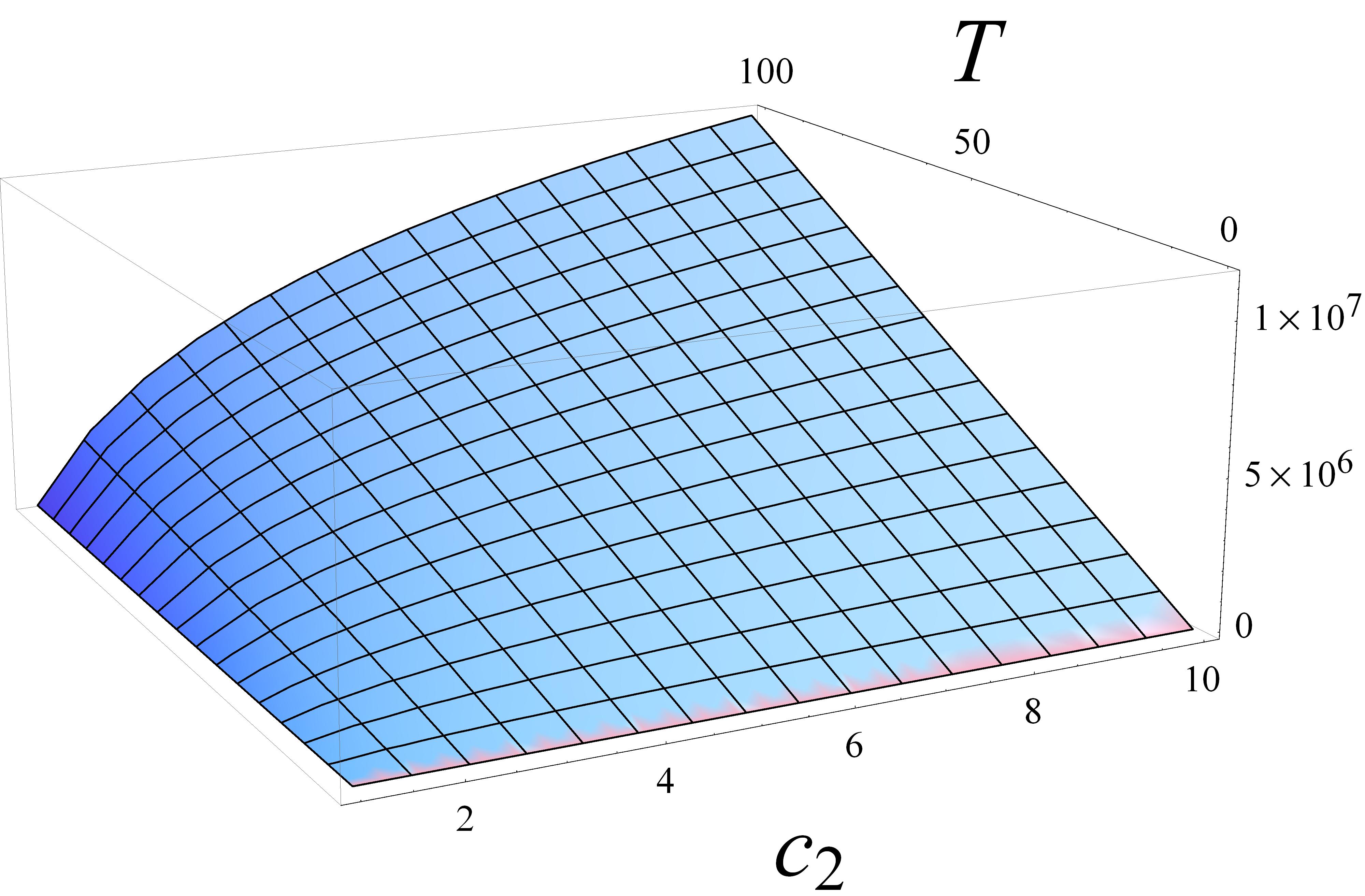}
    \caption{$\frac{1}{M}\frac{1}{V}\frac{dC}{dt}$ in terms of anisotropy parameter $c_2$ and temperature for values of $z=2$ and $\theta=1/2$. }\label{iac}
\end{figure}

To support this idea one can consider IAC which gives rise to the fact that by increase in anisotropy, the critical temperature decreases, then it expects that at a constant temperature, increase in the anisotropy grows the degrees of freedom. This tendency could be seen in Fig(\ref{iac}) that shows the behavior of $\frac{1}{M}\frac{dC}{dt}$ in terms of anisotropy and temperature.

\begin{figure}[htp]
    \centering
    \includegraphics[width=8cm]{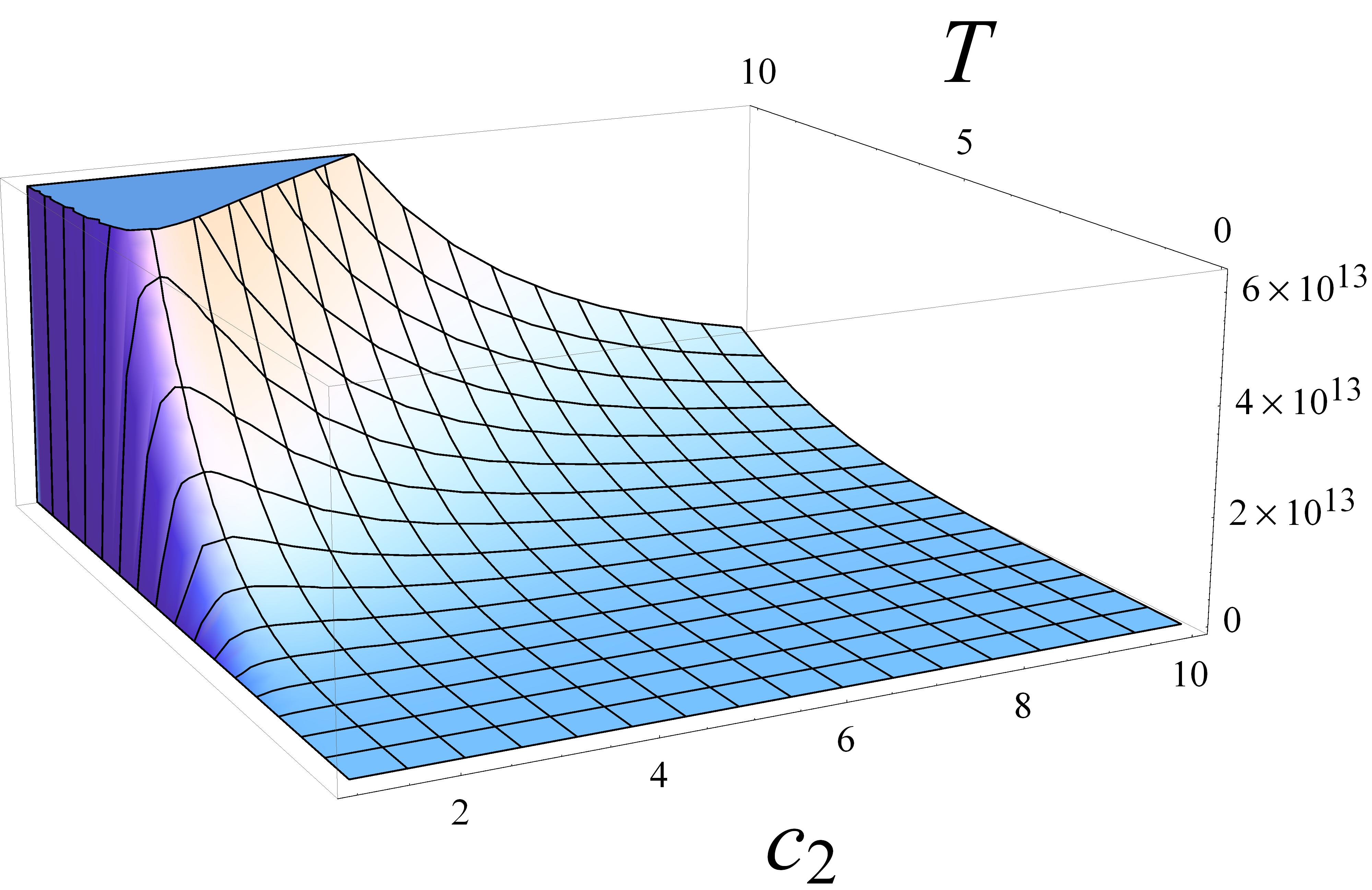}
    \caption{$\frac{1}{V}\frac{dC}{dt}$ in terms of anisotropy parameter $c_2$ and temperature for values of $z=2$ and $\theta=1/2$. }\label{ant}
\end{figure}

Whereas, the complexity growth rate decreases by the increase in the anisotropy at a constant temperature as Fig(\ref{ant}) shows, the degrees of freedom increases which means that the complexity growth rate itself could not be interpreted as the degrees of freedom.

Furthermore, in Vaidya solution we have seen that $\frac{1}{M}\frac{dS_{\text{total}}}{dt}$ is independent of anisotropy then it means that the degrees of freedom in the system do not change when the anisotropy varies and unlike the two-sided black brane solution there is only one behavior of the complexity growth rate. This behavior could be addressed to the lack of phase transition in the one-sided solution. The formation of the black brane is dual to the injection of energy in the boundary and the system from the ground state in the boundary (\eqref{step} with condition $v<0$ in the bulk) goes to an excited state in the boundary with temperature $T$ (\eqref{step} with $v>0$) where the state has different temperature and free energy, as it has been revealed in \cite{2} \footnote{\cite{2} considers quantum phase transition through entanglement entropy; however, thermal phase transitions could be demonstrated in the phase transitions of the entanglement entropy\cite{limitation}.}.

\section{Conclusion}\label{conclusion}
In this work we have studied anisotropic effect on complexity growth rate. The theory of Enstein-Dilaton-Axion has been considered with an Lifshitz anisotropic hyperscaling violation metric and exact dilaton solution in IR limit in term of anisotropy parameter. Because of confinement-deconfinement phases in two-sided black brane there are two different behaviors of the complexity growth rate. In small anisotropy parameters, complexity growth rate respects the Lloyd bound and in large anisotropy parameters it violates. Nonetheless, the Lloyd bound is independent of the anisotropy in the one-sided black brane which is dual to a global quantum quench on the boundary. This stems from the fact that the system does not experience a phase transition under global quantum quench by the injection of energy on the boundary.

By the way, in the extreme case when the anisotropy is large, the complexity growth rate tends to unity which means that there would not be any effort to reach the target state. In Fig(\ref{ani}) it could be seen that near $c_2=1$ call it near isotropy, in two-sided black brane there is a direct relation, in other words, the behavior of the complexity is similar to the case of the isotropic one. Moreover, the Lloyd bound in the near isotropy, just like the isotopic one is respected. Although the same result appears in  \cite{seyed}, their solution is pertuebative in anisotropic parameter and they regarded this effect near the isotropy; however, as anisotropy grows, the contribution of the mass of the black brane dominates and the relation between anisotropy and complexity growth rate would be inversed. Actually the parameter $c_2$ (anisotropy parameter) appears in the mass of the black brane (the coefficient of the complexity growth rate in \eqref{anig}), so the increase in anisotropy shows itself in the mass of the black brane. The physical explanation of this effect is due to role of dilaton filed in string theory and the number of degrees of freedom on the boundary field theory. As anisotropy parameter grows, the value of dilaton filed increases and owing to the fact that dilaton field is in charge of open strings coupling, the open strings would be more coupled which means difficulty of going from reference state to the target would be more severe and as a consequence the complexity growth rate decreases. In \cite{seyed} for the values of anisotropy parameter far away from isotropy, their model shows an conformal anomaly which they expect this leads to the Lloyd bound violation; however, due to the fact that they computed perturbatively they could not show that. In Fig (\ref{10}) the violation has occurred.

It has been suggested that the complexity growth rate itself is not a proper representation of system degrees of freedom but $\frac{1}{M}\frac{dC}{dt}$. This quantity by increase in anisotropy grows which means that according to IAC when the anisotropy parameter increases the critical temperature decreases then it is envisaged at a constant temperature when the anisotropy increases the confinement-deconfinement crossover would lead to a grow in the system degrees of freedom as depicted in Fig(\ref{iac}). However, in the one-sided black brane dual of a global quantum quench there is no phase transition which is due to the energy injection on the boundary. Then as a consequence, the complexity growth rate is independent of the anisotropy.

\section*{Acknowledgment}
The authors would like to thank Juan Pedraza and Mostafa Ghasemi for useful discussions and Hamed Zolfi for careful reading the manuscript. This work is dedicated to Iranian brave women especially M Shahbazi's mother that has just passed away.
\appendix
\renewcommand\theequation{\thesection-\arabic{equation}} 
\setcounter{equation}{0}
\section{Action computation for two-sided black brane}\label{detail}

The contribution of space-like surface $r=r_{max}$ is given:
\begin{equation}
_mn_\mu=\sqrt{\frac{\omega}{f}}r^{\frac{\theta}{dz}-1}dr.
\end{equation}
\begin{align}
K=h^{ab}K_{ab}=h^{ab}n_{\mu}\Gamma^{\mu}_{ba}=\frac{-r^{1-\frac{\theta}{dz}}}{2\sqrt{\omega f}}\Big(f'-\frac{2f}{r}(\frac{d^2z-d\theta+d-\theta}{dz})\Big),
\end{align}
where $_mn^\mu$ is a normal vector to $r=r_{max}$, $a,b$ are integers in $1,... ,d+1$ and $\mu$ in$1,... ,d+2$ . Then GHY term leads to\footnote{In the following we suppose $t_L=t_R=\frac{t}{2}$, owing to the fact that complexity of the system is a function of $t_L+t_R$ and for the matter of convenience we set them as $\frac{t}{2}$. }:
\begin{align}
S_{\text{GHY}}=\frac{V_d}{8\pi G}\sqrt{\frac{\omega^{d}}{\kappa}}\frac{1}{r^{d-1+\frac{1}{z}-\frac{\theta}{z}}}\Big(f'-\frac{2f}{r}(\frac{d^2z-d\theta+d-\theta}{dz})\Big)\Big(\frac{t}{2}+r^*(R)-r^*(r)\Big)\Big|_{r=r_{max}},
\end{align}

\begin{equation}
=\frac{V_d}{8\pi G}\sqrt{\frac{\omega^{d}}{\kappa}}\frac{\Big(d+\frac{1}{z}-\frac{\theta}{z}-\frac{2\theta}{dz}\Big)}{r_h^{d+(\frac{1-\theta}{z})}}\Big(\frac{t}{2}+r^*(R)-r^*(r_{max})\Big). \label{ghy}
\end{equation}
Contribution of GHY term for $r=R$ is time-independent and because of the fact that we are interested in complexity growth rate, it is not needed to compute.

The null boundaries of WDW patch in the bottom of figure (\ref{fig:two}) are given by:

\begin{equation}
t_R-t=r^*(R)-r^*(r),~~~~t+t_L=r^*(L)-r^*(r).
\end{equation}
Then normal vectors to these surfaces are:
\begin{equation}
n_1^\mu=A\frac{r^{2-\frac{2\theta}{dz}}}{\omega}(\frac{1}{f}\partial_t+\partial_r),~~~~~n_3^\mu=B\frac{r^{2-\frac{2\theta}{dz}}}{\omega}(\frac{1}{f}\partial_t-\partial_r),
\end{equation}
where $A$ and $B$ are constant. Contribution of joint points to the action are due to the points $r_m$, $A$ and $B$. In the following $r_m$'s portion is given:
\begin{align}
S_{r_m}=\frac{1}{8\pi G}\int d^dx \sqrt{-\gamma} \log|\frac{n_1.n_3}{2}|,
\end{align}
which leads to:
\begin{equation}
S_{r_m}=\frac{V_d}{8\pi G}\sqrt{\frac{\omega^d}{\kappa}}r^{\frac{\theta-1-dz+z}{z}}\Big((1-\frac{\theta}{dz})\log(r^2)- \log f(r)+\log(\frac{AB}{\omega})\Big)\Big|_{r=r_m}.\label{m}
\end{equation}
Normal vector to the surface $r=R$ would be:
\begin{align}
{n_R}_{\mu}=\sqrt{\frac{\omega}{f}}r^{\frac{\theta}{dz}-1}dr,
\end{align}
and computation of $A$'s portion goes as follows:
\begin{align}
S_{\text{A}}=\frac{\sgn(j)}{8\pi G}\int d^dx \sqrt{-\gamma} \log|n_1.n_R|,
\end{align}
which is written as:
\begin{equation}
S_{\text{A}}=\frac{-V_d}{8\pi G}\sqrt{\frac{\omega^d}{\kappa}} r^{\frac{\theta-1-dz+z}{z}} \log\Big(\frac{A}{\sqrt{\omega f}}r^{1-\frac{\theta}{dz}}\Big)\Big|_{r=R}.\label{a}
\end{equation}
The same calculation for joint point $B$ occurs:
\begin{equation}
S_{\text{B}}=\frac{-V_d}{8\pi G}\sqrt{\frac{\omega^d}{\kappa}} r^{\frac{\theta-1-dz+z}{z}} \log\Big(\frac{D}{\sqrt{\omega f}}r^{1-\frac{\theta}{dz}}\Big)\Big|_{r=R}, \label{b}
\end{equation}
where $D$ is the normalization coefficient for normal vector to surface $2$. The joint point contribution for points $A$ and $B$\footnote{For their counterparts on the left side of figure (\ref{fig:two}) are time-independent too.} are time-independent.

To eliminate ambiguities arises by normalization of null vectors we should add some counter terms which are written as follows \cite{mya}:
\begin{equation}
S_{\text{null sur.}}=\frac{-1}{8\pi G}\int d\tau d^dx \sqrt{-\gamma} \Theta \log \frac{\Theta}{d-\theta},
\end{equation}
where for term $d-\theta$ we followed \cite{alish} and $\Theta$ could be written as:
\begin{align}
\Theta=\frac{1}{\sqrt{\gamma}}\frac{\partial \sqrt{\gamma}}{\partial \tau}.
\end{align}
For surface $1$, $\Theta$ is given by:
\begin{align}
\Theta=\frac{A}{\omega z}(\theta-1-dz+z) r^{1-\frac{2\theta}{dz}},
\end{align}
so the counter term for surface $1$ would be:
\begin{equation}
S_{\text{null sur.}}^1=\frac{V_d}{8\pi G}\sqrt{\frac{\omega^d}{\kappa}}\frac{z r^{\frac{\theta-1-dz+z}{z}}}{\theta-1-dz+z} \bigg((1-\frac{2\theta}{dz})+(\frac{dz+1-\theta-z}{z})\log \Big(\frac{A}{\omega z(d-\theta)}(\theta-1-dz+z)r^{1-\frac{2\theta}{dz}}\Big)\bigg)\Big|_{r=R}^{r=r_m}. \label{ns}
\end{equation}
The same expression is repeated for surface $3$ but the boundary would be from $r=L$\footnote{It is supposed that the value of $r=R$ and $r=L$ are the same.} to $r=r_m$, we should note that only terms containing $r=r_m$ are time-dependent. One can show that the similar behavior is shown up for surface $2$ and $4$ but by replacement of $r_m$ for $r_{max}$.
Furthermore, we should add some counter terms for time-like surfaces which are read as \cite{sken}:
\begin{equation}
S_{\text{time-like}}=-\frac{1}{16\pi G}\int d^dx dt\sqrt{-h}\Big(2d+\frac{R}{d}+O(R^2)\Big),
\end{equation}
where $h$ is induced metric and $R$ is Riemann curvature of $h$. For surface $r=R$, it is read as:
\begin{equation}
S_{\text{time-like}}=\frac{-V_d \Delta t}{16\pi G} \sqrt{\frac{\omega^{d+1}}{\kappa}}\frac{\sqrt{f}}{r^{d+\frac{1}{z}-\frac{(d+1)\theta}{dz}}}\Big|_{r=R},\label{time}
\end{equation}
where $\Delta t$ is independent of time and is given by:
\begin{align}
\Delta t=2\int_0^{r=R}\frac{dr}{f},
\end{align}
then eq. \eqref{time} has no contribution to complexity growth rate. There is one step remained to have complexity growth rate and that is computing eq. \eqref{reform} in WDW patch. There are four regions that their contributions would be as follows:
\begin{align}
&S_{\text{bulk}}^1=-\frac{6V_d}{4\pi Gd}\sqrt{\frac{\omega^{d+2}}{\kappa}}e^{\tau c_3} c_2^{\tau c_1} \int_{r_h}^{r_{max}}  ~ r^{\frac{(d+2)\theta}{dz}-\frac{1}{z}-(d+1)+\tau c_1}\Big(\frac{t}{2}+r^*(0)-r^*(r)\Big) dr,\\
&S_{\text{bulk}}^2=S_{\text{bulk}}^4=-\frac{6V_d}{4\pi Gd}\sqrt{\frac{\omega^{d+2}}{\kappa}}e^{\tau c_3} c_2^{\tau c_1} \int_{r=R}^{r_{h}}  ~ r^{\frac{(d+2)\theta}{dz}-\frac{1}{z}-(d+1)+\tau c_1}\Big(r^*(0)-r^*(r)\Big) dr,\\
&S_{\text{bulk}}^3=-\frac{6V_d}{4\pi Gd}\sqrt{\frac{\omega^{d+2}}{\kappa}}e^{\tau c_3} c_2^{\tau c_1} \int_{r_h}^{r_{m}}  ~ r^{\frac{(d+2)\theta}{dz}-\frac{1}{z}-(d+1)+\tau c_1}\Big(-\frac{t}{2}+r^*(0)-r^*(r)\Big) dr.
\end{align}
The contribution of the whole bulk could be written in terms of time-dependent and independent terms:

\begin{equation}
\begin{aligned}
S_{\text{bulk}}=&-\frac{6V_d}{4\pi Gd}\sqrt{\frac{\omega^{d+2}}{\kappa}}e^{\tau c_3} c_2^{\tau c_1} \int_{r_m}^{r_{max}}  ~ r^{\frac{(d+2)\theta}{dz}-\frac{1}{z}-(d+1)+\tau c_1}\Big(\frac{t}{2}-r^*(0)+r^*(r)\Big) dr \\
&-\frac{3V_d}{\pi Gd}\sqrt{\frac{\omega^{d+2}}{\kappa}}e^{\tau c_3} c_2^{\tau c_1} \int_{r=R}^{r_{max}}  ~ r^{\frac{(d+2)\theta}{dz}-\frac{1}{z}-(d+1)+\tau c_1}\Big(r^*(0)-r^*(r)\Big) dr. \label{bulk}
\end{aligned}
\end{equation}

It is still needed to add a counter term to engender the action finite which could be written as \cite{alish}:
\begin{align}
S_{\text{count.}}=\frac{1}{8\pi G}\int d\tau d^dx \sqrt{\gamma} \Theta \Big(\frac{1}{2}\alpha \phi+L\Big),
\end{align}
which for light-like surface 1 leads to:
\begin{equation}
S_{\text{count.}}=-\frac{V_d}{8\pi G} \sqrt{\frac{\omega^d}{\kappa}} r^{\frac{\theta-dz-1+z}{z}}\bigg(\frac{\alpha c_1}{2}\Big(\frac{z}{z+\theta-1-dz}+\log(c_2r)\Big)+\frac{\alpha c_3}{2}+L\bigg)\Big|_{r=R}^{r=r_{m}}. \label{count}
\end{equation}
The coefficients $\alpha$ and $L$ are read as:
\begin{align}
\alpha c_1=\frac{2\theta}{dz},
\end{align}

\begin{align}
2L=\log(\omega f(R)) -\frac{2z(1-\frac{2\theta}{dz})}{dz+1-z-\theta}+\frac{2\theta}{d(z+\theta-1-dz)}-\alpha c_3+2\log(\frac{\theta-1-dz+z}{\omega z(d-\theta)})+\frac{2\theta}{dz}\log c_2.
\end{align}

Now we are able to compute the rate of complexity by total action from eq.s \eqref{ghy} \eqref{m} \eqref{a} \eqref{b} \eqref{ns} \eqref{time} \eqref{bulk} \eqref{count}:
\begin{equation}
\begin{aligned}
S_{\text{total}}=&S_{\text{GHY}}+S_{\text{$r_m$}}+S_{\text{A}}+S_{\text{B}}+S_{\text{null sur.}}^1+S_{\text{null sur.}}^2+S_{\text{null sur.}}^3+S_{\text{null sur.}}^4\\
&+S_{\text{time-like}}+S_{\text{bulk}}+S_{\text{count.}}.
\end{aligned}
\end{equation}
The leading term of the complexity in large anisotropy arises by counter terms \eqref{count}:
\begin{align} 
S_{\text{total}}\propto \sqrt{\frac{\omega^d}{\kappa}} \log(c_2)=c_2^{-d+1\frac{\theta-1}{z}}\log(c_2),
\end{align}\label{complexity}
which is decreasing by growing the anisotropy.
\section{One-sided black brane}\label{vaidyaa}
It is needed to consider the patches $v>0$ and $v<0$.
\\
\\
\textbf{\large $v>0$}
\\

For $v>0$ there are five boundaries: $v=0$, $v=t$, $r=R$, $r=r_m$ and line going through $BP$ which is given by:
\begin{equation}
t-v=2\int _R ^{r(v)} \frac{dr}{f}.
\end{equation}
Normal vector to surface $r=r_{max}$ would be:
\begin{align}
n_m^{\mu}=\frac{r^{1-\frac{\theta}{dz}}}{\sqrt{\omega}}\Big(-\frac{1}{\sqrt{f}}\partial_v+\sqrt{f}\partial_r\Big),
\end{align}
and extrinsic curvature reads as:
\begin{align}
K=h^{ab}K_{ab}=-\sqrt{\frac{\omega}{f}}\frac{ r^{1-\frac{\theta}{dz}}}{2\omega}\bigg(f'-\frac{2f}{r}\Big(\frac{d^2z-d\theta +d-\theta}{dz}\Big)\bigg).
\end{align}
Put these all in eq. \eqref{ghy}:
\begin{equation}
\begin{aligned}
S_{\text{GHY}}=&\frac{V_d}{16\pi G}\sqrt{\frac{\omega^d}{\kappa}} r^{\frac{\theta-1}{z}+1-d}\bigg(f'-\frac{2f}{r}\Big(\frac{d^2z-d\theta+d-\theta}{dz}\Big)\bigg)t\Big|_{r=r_{max}}\\
&=\frac{V_d}{16\pi G}\sqrt{\frac{\omega^d}{\kappa}} \frac{t}{r_h^{d+\frac{1-\theta}{z}}}\Big(\frac{dz-\theta+1-\frac{2\theta}{d}}{z}\Big).
\end{aligned}
\end{equation}

For joint points, normal vector to the surfaces is needed. Normal vectors to $v=0 (v=t)$, $BP$ and $r=R$ respectively are given:
\begin{align}
&n_0^{\mu}=-\frac{B r^{2-\frac{2\theta}{dz}}}{\omega}\partial_r\\
&n_{BP}^{\mu}=\frac{C r^{2-\frac{2\theta}{dz}}}{\omega}\Big(-\frac{2}{f}\partial_v+\partial_r\Big)\\
&n_R^{\mu}=\frac{r^{1-\frac{\theta}{dz}}}{\sqrt{\omega}}\Big(-\frac{1}{\sqrt{f}}\partial_v+\sqrt{f}\partial_r\Big).
\end{align}
Then, the contribution of joint points for point $P$ would be:
\begin{equation}
S^{v>0}_{\text{P}}=\frac{V_d}{8\pi G}\sqrt{\frac{\omega^d}{\kappa}}r^{\frac{\theta-1}{z}+1-d}\bigg((1-\frac{\theta}{dz}) \log r^2 -\log f(r) +\log \frac{BC}{\omega}\bigg)\Big|_{r=r_P}.
\end{equation}
The contribution of points $A$ and $B$ would be similar which is given by:
\begin{equation}
S_{\text{A}}=\frac{V_d}{8\pi G}\sqrt{\frac{\omega^d}{\kappa}}r^{\frac{\theta-1}{z}+1-d}\bigg((1-\frac{\theta}{dz}) \log r -\log \sqrt{f(r)} +\log \frac{B}{\sqrt{\omega}}\bigg)\Big|_{r=R},
\end{equation}
which is time-independent, then the joint point contribution for point $A$ and $B$ are negligible for complexity growth rate. 
The bulk action contribution to complexity as shown in eq. \eqref{reform} goes as follows:
\begin{align}
S_{\text{bulk}}=-\frac{3V_d}{4\pi Gd}\sqrt{\frac{\omega^{d+2}}{\kappa}} e^{\tau c_3} c_2^{\tau c_1} \int dv \int dr~ r^{\frac{(d+2)\theta}{dz}-\frac{1}{z}-d-1+\tau c_1}.
\end{align}
For region $v>0$, it leads to:
\begin{equation}
\begin{aligned}
S^{v>0}_{\text{bulk}}=&-\frac{3V_d}{4\pi Gd}\sqrt{\frac{\omega^{d+2}}{\kappa}} e^{\tau c_3} c_2^{\tau c_1}\Big\{ \int_0^{v_B} dv \int_{r(v)}^{\infty} dr ~r^{\frac{(d+2)\theta}{dz}-\frac{1}{z}-d-1+\tau c_1}\\
&+\int_{v_B}^t dv \int_{r_B=R}^{\infty} dr ~r^{\frac{(d+2)\theta}{dz}-\frac{1}{z}-d-1+\tau c_1}\Big\},
\end{aligned}
\end{equation}
where $v_B\approx t-2r_B$ . By assumption $\frac{(d+2)\theta}{dz}-\frac{1}{z}-d-1+\tau c_1<-1$, one can get to the following result:
\begin{equation}
\begin{aligned}
S^{v>0}_{\text{bulk}}=&\frac{3V_d}{4\pi Gd}\sqrt{\frac{\omega^{d+2}}{\kappa}} \frac{ e^{\tau c_3} c_2^{\tau c_1}}{\Big(\frac{(d+2)\theta}{dz}-\frac{1}{z}-d+\tau c_1\Big)}\Big\{\int_0^{v_B=t-2r_B} dv ~r(v)^{\frac{(d+2)\theta}{dz}-\frac{1}{z}-d+\tau c_1}+2 r_B^{\frac{(d+2)\theta}{dz}-\frac{1}{z}-d+1+\tau c_1}\Big\}\\
&=\frac{6V_d}{4\pi Gd}\sqrt{\frac{\omega^{d+2}}{\kappa}} \frac{ e^{\tau c_3} c_2^{\tau c_1}}{\Big(\frac{(d+2)\theta}{dz}-\frac{1}{z}-d+\tau c_1\Big)}\Big\{\frac{r_P^{\frac{(d+2)\theta}{dz}-\frac{1}{z}-d+1+\tau c_1}}{\Big(\frac{(d+2)\theta}{dz}-\frac{1}{z}-d+1+\tau c_1\Big)}+r_B^{\frac{(d+2)\theta}{dz}-\frac{1}{z}-d+1+\tau c_1}\Big\}.
\end{aligned}
\end{equation}

\textbf{\large $v<0$}
\\

In this region the function $f$ is 1 and the work is easier. The contribution of joint point $P$ is similar to its counterpart in $v>0$ but for $f=1$ and a minus, so it is written as:

\begin{equation}
S^{v<0}_{\text{P}}=\frac{-V_d}{8\pi G}\sqrt{\frac{\omega^d}{\kappa}}r^{\frac{\theta-1}{z}+1-d}\bigg((1-\frac{\theta}{dz}) \log r^2 +\log \frac{BC}{\omega}\bigg)\Big|_{r=r_P}.
\end{equation}
Finally, one can compute the bulk contribution of complexity in the region $v<0$:
\begin{equation}
\begin{aligned}
S_{\text{bulk}}^{v<0}&=-\frac{3V_d}{4\pi Gd} \sqrt{\frac{\omega^{d+2}}{\kappa}} e^{\tau c_3} c_2^{\tau c_1} \int_{-\infty}^0 dv \int_{r_P-\frac{v}{2}}~r^{\frac{(d+2)\theta}{dz}-\frac{1}{z}-d-1+\tau c_1}\\
&=\frac{-6V_d}{4\pi Gd}\sqrt{\frac{\omega^{d+2}}{\kappa}} e^{\tau c_3} c_2^{\tau c_1} \frac{r_P^{\frac{(d+2)\theta}{dz}-\frac{1}{z}-d+1+\tau c_1}}{\Big(\frac{(d+2)\theta}{dz}-\frac{1}{z}-d+\tau c_1\Big)\Big(\frac{(d+2)\theta}{dz}-\frac{1}{z}-d+1+\tau c_1\Big)}.
\end{aligned}
\end{equation}



\begin{thebibliography}{99}

\bibitem{mal} J. M. Maldacena, ''The Large N Limit of Superconformal Field Theories and Supergravity,'' Adv.Theor.Math.Phys.2:231-252,1998, arXiv:9711200v3[hep-th].
\bibitem{comp} L. Susskind, "Three Lectures on Complexity and Black Holes,"	arXiv:1810.11563[hep-th].
\bibitem{shock} L. Susskind and D. Stanford, "Complexity and shock wave geometry," Phys. Rev. D 90, 126007 (2014), arXiv:1406.2678[hep-th].
\bibitem{ca} A. R. Brown, D. A. Roberts, L. Susskind, B. Swingle, and Y. Zhao, "Complexity Equals Action," Phys. Rev. Lett. 116, 191301 (2016), 	arXiv:1509.07876 [hep-th].
\bibitem{couch}J. C., W. Fischler and P. H. Nguyen, ''Noether charge, black hole volume, and complexity," JHEP 1703 (2017) 119, arXiv:1610.02038[hep-th].
\bibitem{anything}A. Belin, R. C. Myers, S.-M. Ruan, G. Sárosi, and A. J. Speranza, "Does Complexity Equal Anything?," 
Phys.Rev.Lett. 128 (2022) 8, 081602, arXiv: 2111.02429 [hep-th]
\bibitem{lloyd} S. Lloyd, "Ultimate physical limits to computation," Nature 406 (Aug., 2000), arXiv:9908043[quant-ph].

\bibitem{myers} D. Carmi, S. Chapman, H. Marrochio, R. C. Myers, and S. Sugishita, "On the Time Dependence of Holographic Complexity," JHEP 1711 (2017) 188, arXiv:1709.10184 [hep-th].
\bibitem{l1} M. Moosa, "Divergences in the rate of complexitication," Phys. Rev. D 97, 106016 (2018), arXiv:1712.07137[hep-th].
\bibitem{l2} M. Ghodrati, "Complexity growth in massive gravity theories, the effects of chirality, and more," Phys. Rev. D 96, no. 10, 106020 (2017).
\bibitem{l3} R. Q. Yang, C. Niu, C. Y. Zhang, and K. Y. Kim, "Comparison of holographic and field theoretic complexities for time-dependent thermofield double states," JHEP 1802, 082 (2018).
\bibitem{zolfi}H. Zolfi, "Complexity and Multi-boundary Wormholes in 2 + 1 dimensions," JHEP 04 (2023) 076, arXiv:2302.07522 [hep-th].

\bibitem{alish} M. Alishahiha, A. Faraji Astaneh, M. R. Mohammadi Mozaffar, and A. Mollabashi, "Complexity Growth with Lifshitz Scaling and Hyperscaling Violation," JHEP 1807 (2018).

\bibitem{mod1}R.-Q. Yang, C. Niu, C.-Y. Zhang and K.-Y. Kim, "Comparison of holographic and field theoretic complexities by time dependent thermofield double states," JHEP 02 (2018) 082, arXiv:1710.00600v2 [hep-th].
\bibitem{mod2}A. Akhavan, M. Alishahiha, A. Naseh and H. Zolfi, "Complexity and behind the horizon cut off", JHEP12(2018)090.
\bibitem{mod3}J. Couch, S. Eccles, T. Jacobson and P. Nguyen, "Holographic Complexity and Volume," JHEP 11 (2018) 044, arXiv:1807.02186v2 [hep-th].

\bibitem{e1} A. Karch and A. O'Bannon, ''Metallic AdS/CFT," JHEP 09 (2007) 024 [0705.3870].
\bibitem{e11}T. Albash, V. G. Filev, C. V. Johnson, and A. Kundu,''Quarks in an external electric field in
finite temperature large N gauge theory," JHEP 08 (2008) 092 [0709.1554].
\bibitem{e12}E. D'Hoker and P. Kraus,''Magnetic Brane Solutions in AdS," JHEP 10 (2009) 088, arXiv:0908.3875[hep-th].
\bibitem{e13}K. Jensen, A. Karch, and E. G. Thompson, ''A Holographic Quantum Critical Point at Finite
Magnetic Field and Finite Density," JHEP 05 (2010) 015, arXiv:1002.2447[hep-th].
\bibitem{e14}K.-Y. Kim, B. Sahoo, and H.-U. Yee, ''Holographic chiral magnetic spiral," JHEP 10 (2010)
005, arXiv:1007.1985[hep-th].
\bibitem{e15}N. Evans, K.-Y. Kim, J. P. Shock, and J. P. Shock, ''Chiral phase transitions and quantum
critical points of the D3/D7(D5) system with mutually perpendicular E and B fields at finite
temperature and density," JHEP 09 (2011) 021, arXiv:1107.5053[hep-th].
\bibitem{e16}A. Donos, J. P. Gauntlett, and C. Pantelidou, ''Magnetic and Electric AdS Solutions in
String- and M-Theory," Class. Quant. Grav. 29 (2012) 194006, arXiv:1112.4195[hep-th].



\bibitem{e2}U. Gursoy, M. Jarvinen, G. Nijs, and J. F. Pedraza, ''Inverse Anisotropic Catalysis in Holographic QCD," JHEP 04 (2019) 071,  arXiv:1811.11724v3.

\bibitem{d1}D. Mateos and D. Trancanelli, ''Thermodynamics and Instabilities of a Strongly Coupled
Anisotropic Plasma," JHEP 07 (2011) 054, arXiv:1106.1637[hep-th].
\bibitem{d11}P. Liu, C. Niu, and J.-P. Wu, ''The Effect of Anisotropy on Holographic Entanglement Entropy
and Mutual Information," Phys. Lett. B796 (2019) 155, arXiv:1905.06808[hep-th].
\bibitem{d12}D. Roychowdhury, ''Holography for anisotropic branes with hyperscaling violation," JHEP 01
(2016) 105.
\bibitem{d13}A. Donos, J. P. Gauntlett, and O. Sosa-Rodriguez, ''Anisotropic plasmas from axion and
dilaton deformations," JHEP 11 (2016) 002.
\bibitem{strong} D. Giataganas, U. Grsoy, and J. F. Pedraza, "Strongly-coupled anisotropic gauge theories and holography," Phys. Rev. Lett. 121, no. 12, 121601 (2018), arXiv:1708.05691.


\bibitem{both}D. M. Rodrigues, E. F. Capossoli, and H. Boschi-Filho, "Magnetic catalysis and inverse magnetic catalysis in (2+1)-dimensional gauge theories from holographic models," Phys. Rev. D 97, 126001.
\bibitem{interply}U. Gürsoy, M. Järvinen, G. Nijs, and J. F. Pedraza, "On the interplay between magnetic field and anisotropy in holographic QCD," JHEP 2103 (2021) 180, arXiv:2011.09474v3 [hep-th].


\bibitem{EDA1}M. J. Vasli, M. R. Mohammadi Mozaffar, K. Babaei Velni and M. Sahraei, "Holographic Study of Reflected Entropy in Anisotropic Theories," Phys.Rev.D 107 (2023) 2, 026012, arXiv:2207.14169 [hep-th].
\bibitem{2}P.-C. Sun, D.-S. Lee and C.-P. Yeh, "Holographic approach to thermalization in general anisotropic theories," JHEP 03 (2021) 164, arXiv:2011.02716v2 [hep-th].
\bibitem{3}M. Ghasemi and S. Parvizi, "Constraints on anisotropic RG flows from holographic entanglement entropy," Phys.Rev.D 104 (2021) 086028, arXiv:1907.01546v3 [hep-th].
\bibitem{EDA2}M. Ghasemi and S. Parvizi, "Curved Corner Contribution to the Entanglement Entropy in an Anisotropic Spacetime,"  arXiv:1905.01675v4 [hep-th].






\bibitem{string} D. Mateos and D. Trancanelli, ''Thermodynamics and Instabilities of a Strongly Coupled Anisotropic Plasma," JHEP 1107, 054 (2011),
arXiv:1106.1637 [hep-th].


\bibitem{pad} K. Parattu, S. Chakraborty, B. R. Majhi, and T. Padmanabhan, "A Boundary Term for the Gravitational Action with Null Boundaries," Gen. Rel. Grav. 48, no. 7, 94 (2016), arXiv:1501.01053.
\bibitem{myersnul}L. Lehner, R. C. Myers, E. Poisson, and R. D. Sorkin, "Gravitational action with null boundaries," Phys. Rev. D 94 (2016) no.8, 084046, arXiv:1609.00207.
\bibitem{mya}L. Lehner, R. C. Myers, E. Poisson, and R. D. Sorkin, "Gravitational action with null boundaries," Phys. Rev. D 94 (2016) no.8, 084046, arXiv:1609.00207 [hep-th].
\bibitem{sken} M. Henningson and K. Skenderis, "The Holographic Weyl anomaly," JHEP 9807, 023 (1998), arXiv:9806087.
\bibitem{ghodrati}M. Ghodrati, "Complexity growth rate during phase transitions," PHYSICAL REVIEW D 98, 106011 (2018).
\bibitem{dil}Y.-S. An and R.-H. Peng, "The effect of dilaton on the holographic complexity growth,"  Phys. Rev. D 97, 066022 (2018).

\bibitem {seyed} S. A. Hosseini Mansoori, V. Jahnke, M. M. Qaemmaqami, and Y. D. Olivas, "Holographic complexity of anisotropic black branes," Phys. Rev. D 100, 046014 (2019), arXiv:1808.00067v3.
\bibitem{gri}D. J. Griffiths, "Introduction to Quantum Mechanics (2nd Edition)," Pearson Prentice Hall, 2004.


\bibitem{mosa}M. Moosa, "Evolution of Complexity Following a Global Quench," arXiv:1711.02668v3 [hep-th].

\bibitem{random}P. Saad, S. H. Shenker, and D. Stanford, “JT gravity as a matrix integral,”
arXiv:1903.11115 [hep-th].

\bibitem{averaging}Vijay Balasubramaniana, Matthew DeCrossa, Arjun Kara and Onkar Parrikar, "Quantum
Complexity of Time Evolution with Chaotic Hamiltonians," JHEP 01 (2020) 134 ,arXiv:1905.05765v3 [hepth]

\bibitem{qcd}S.-J. Zhang, ''Complexity and phase transitions in a holographic QCD model," Nucl.Phys. B929 (2018) 243-253,  arXiv:1712.07583 [hep-th].

\bibitem{limitation}N. Jokela, H. Ruotsalainen and J. G. Subils "Limitations of entanglement entropy in detecting thermal phase transitions,"  arXiv:2310.11205v2 [hep-th].

\bibitem{aref}I. Y. Aref'eva, "QGP time formation in holographic shock waves model of heavy ion collisions," Theor.Math.Phys. 184 (2015), 1239-1255, arXiv:1503.02185v1 [hep-th].



\end{thebibliography}
\end{document}